\documentclass{eas}
\usepackage{graphicx}
\pdfoutput=1 
\begin{document}

\title{Magnetic fields and the location of the PDR}
\author{Gary J. Ferland}\address{Institute of Astronomy, Madingley Road,Cambridge CB3 0HA, UK}\address{Department of Physics and Astronomy, University of Kentucky, Lexington KY 40506 USA}
\begin{abstract}
I review recent studies of the emission-line regions
in Orion and M17.
Both have similar geometries, a bubble of hot shocked gas surrounding
the central star cluster, with H$^+$, H$^0$, and H$_2$ regions,
often referred to as H~II regions, PDRs, and molecular clouds,
forming successive shells on the surface of a molecular cloud.
The magnetic fields in the H$^0$ regions have been measured with 21 cm
Zeeman polarization and are found to be 1 -- 2 dex stronger
than the field in the diffuse ISM.
The regions appear to be in rough hydrostatic equilibrium.
The H$^+$ region is pushed away from the star cluster
by starlight radiation pressure.
Since most starlight is in ionizing radiation, most of its outward
push will act on the H$^+$ region and then on to the H$^0$ region.
The magnetic pressure in the H$^0$ region balances the momentum in starlight
and together they set the location of the H$^0$ region.
The picture is that, when the star cluster formed, it created a bubble of ionized gas
which expanded and compressing surrounding H$^0$ and H$_2$ regions.
The magnetic field was amplified until its pressure
was able to support the momentum in starlight.
This offers a great simplification in understanding the underlying
physics that establishes parameters for PDR models.
\end{abstract}
\maketitle
\section{Introduction}

The emission-line regions surrounding newly formed stars share a common geometry.
Hot stars are too short lived to stray far from their birthplace.
As a result the radiation field emitted by a star cluster will strike
the nearby molecular cloud, producing ablating layers where hydrogen is
successively H$^+$, H$^0$, and H$_2$.
Dynamical interactions establish the location of these layers while
both the distance and extinction between the stars and a parcel
of gas establishes its chemical, ionization, and thermal state,
and hence the spectrum that the gas emits.
The technology to study the H$^+$, H$^0$, and H$_2$
regions became available at different times
and, as a result, they have different names, H~II region, PDR, and molecular
cloud respectively.
These regions are actually a continuous flow with
conditions in the gas changing is a systematic manner.
With modern observational facilities all can be studied on an equal basis.

The central question addressed here is the pressure source that sets the structure of
the gas.
Baldwin et al. (1991) built a model of the Orion H$^+$ region, reproducing the optical
and IR emission, by assuming that the ionized gas is in hydrostatic equilibrium.
Their model balanced the outward momentum in starlight with internal pressures
including gas pressure and internally generated line radiation pressure.
The pressure, in commonly used $nT$ units, was $\sim 10^{7} \rm{cm}^{-3}$~K.
That model stopped near the H$^+ - \rm{H}^0$ ionization front and did not
attempt to reproduce lines from atomic or molecular regions.

The H$^+$, H$^0$, and H$_2$ regions are coupled by dynamics.
If gas pressure remains constant across the flow then the densities
would decrease to compensate for the
increase in temperature as the gas moves from neutral regions, with
$T \sim 10^2 - 10^3$~K, to the H$^+$ region with $T\sim 10^4$~K.
The ratio of densities to either side of the I-front would 
reach $n^0/n^+ \sim 10^2$.
In reality the gas is accelerated to roughly the sound speed as it passes through
the ionization front and half of this increase goes into ram pressure (Henney et al. 2005).

Magnetic fields add a further complication.
Flux freezing couples the field to even weakly ionized gas.   Expansion along
field lines will not affect $B$ but motion across field lines will.
The simplest case is a disorganized field, where $B \propto n^\gamma$,
The magnetic pressure, $B^2/8\pi$, is proportional to $n^2$, which
means that, while gas pressure may dominate in the H$^+$ region, the field
will be considerably more important in the denser H$^0$ region.

Magnetic fields, turbulent motions, and cosmic rays are observed to be in energy
equipartition in the general ISM (Lequeux 2004).
It has long been known
that line widths in the H$^0$ region suggest motions corresponding
to a turbulent pressure that is greater than the gas pressures
(Tielens \& Hollenbach 1985; Ferland 2001, Roshi 2007).
It should not be surprising if magnetic fields and cosmic rays were also
energetically important.

Figure 1 shows sketches of the Orion and M17 star-forming regions.
Maps of the magnetic field have been made by measuring 21 cm absorption
of the continuum produced in the H$^+$ region.
The magnetic field in the H$^0$ region causes a small velocity shift in the
left and right hand circularly polarized 21 absorption line.  This shift is
proportional to the line-of-sight component of the $B$ field.
The opacity of the 21 cm line is proportional to $n(H^0)/T_{spin}$
so the measurement is weighted to cold atomic regions.
In Orion the field is measured (Troland et al.\ 1989) in the ''Veil'',
the atomic layers towards
the bottom of Figure 1, while in M17 measurements (Brogan et al.\ 1999)
are of the atomic layer that is
between the H$^+$ region and foreground molecular gas.
Both fields approach half a mG, nearly 2 dex larger than
the diffuse ISM magnetic field.
For reference a 100 $\mu$G field corresponding to a pressure of
\begin{eqnarray}
P_B /k \approx \frac{{B^2 }}{{8\pi k}} \approx B_{100\,\mu G}^2 2.9 \times 10^6
\ \rm{cm}^{-3}\ \rm{K}
\end{eqnarray}

\begin{figure}
\protect\resizebox{\columnwidth}{!}
{\includegraphics{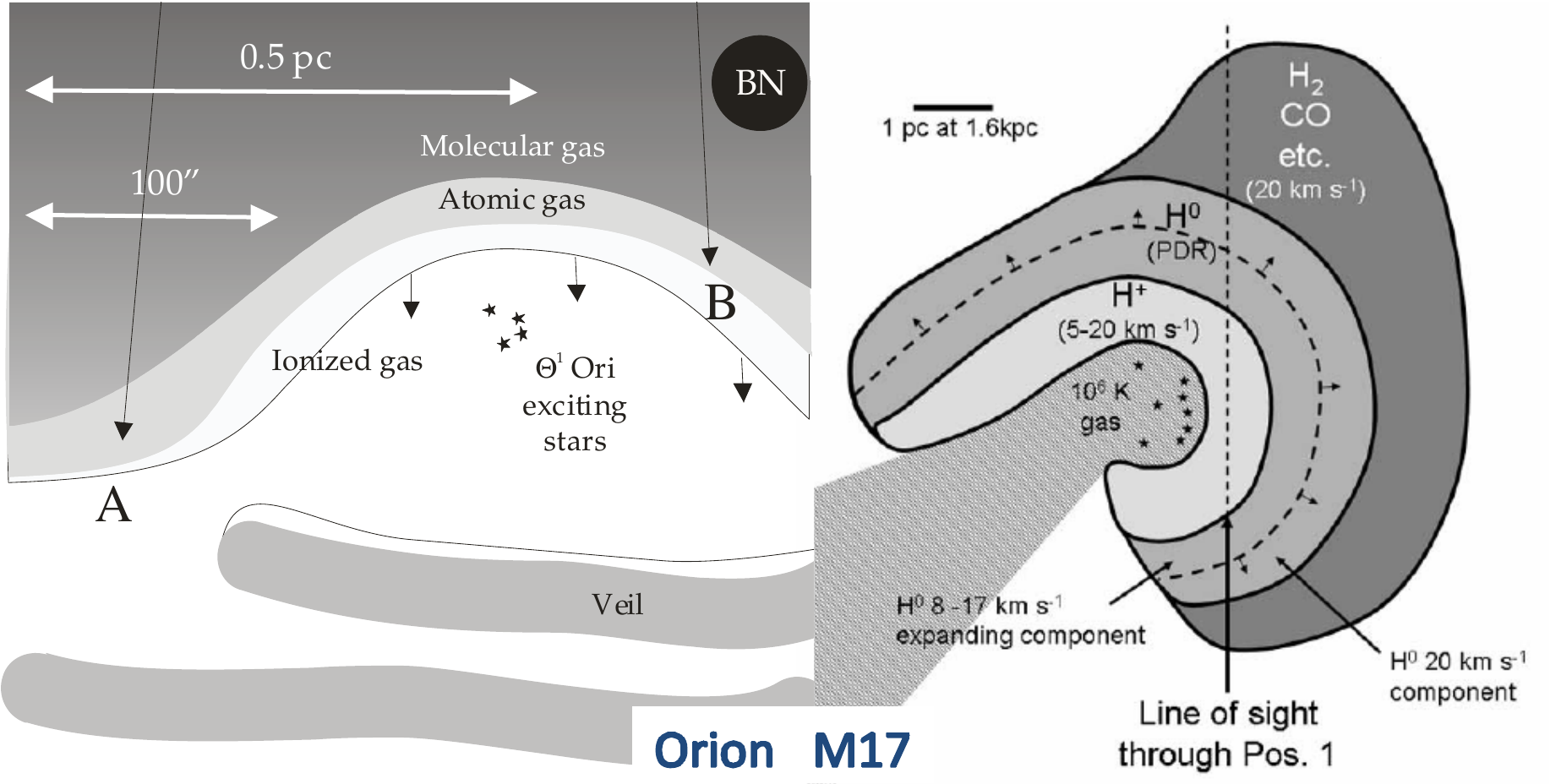}}
\caption{The left panel (from Osterbrock \& Ferland 2006) shows the Orion environment
and the right panel shows M17 (Pellegrini et al. 2007).  The Earth is to the
bottom and the H$^+$, H$^0$, and H$_2$ regions are shown
with lighter to darker shading.
These two objects form a complementary pair, with the molecular cloud directly behind the
star cluster in Orion and off to the right in M18.
We view directly into the flow in Orion while in M17 we view it in profile.
}
\label{fig:temperature}
\end{figure}

The fields in these regions are among the strongest measured
in diffuse matter.
What is the relationship between the magnetic and gas pressures?
How can such high fields be produced and maintained?

Abel et al. (2004, 2006) used UV absorption lines to determine $T$ and $n_{\rm{H}}$
in the atomic layers where $B$ is measured.
They found $P_{gas}/P_{mag} \sim 10$ and that the field and turbulence were not in equipartition.
Their resulting model reproduced the column densities of many atoms, ions, and
ro-vib excited H$_2$.

Pellegrini et al. (2007, hereafter P07)
obtained very deep optical spectra of the heavily obscured
H$^+$ region in M17 to measure the gas pressure in a layer adjacent to
where $B$ is known.
They produced a hydrostatic equilibrium model which reproduced the measured $B$,
the extent of the H$^+$, H$^0$, and H$_2$ regions, and the observed emission
in a range of atomic and ionized species.
The M17 geometry was fully observationally constrained with the cosmic ray density
the only free parameter.
A cosmic ray density $\sim 300$ stronger than the galactic background,
but well below the value for equipartition with $B$,
provided enough heating to account for atomic lines from the H$^0$ region.

What accounts for these strong magnetic fields?
The magnetic pressure balances the total momentum in the stellar radiation field (P07).
A simple picture emerges.
Quiescent molecular clouds have a magnetic field that is related to other properties
(Myers \& Goodman 1998).
When stars form their light pushes surrounding gas away.
The bulk of the radiative momentum is in ionizing radiation so much of this push
acts on the H$^+$ region, the layer where
most ionizing radiation is absorbed.
This outward push is then exerted against the H$^0$ region.
The atomic gas is compressed and the field amplified.
The magnetic pressure increases more rapidly than the gas pressure
and eventually the magnetic pressure balances the starlight.
Algebraically this balance can be expressed as:

\begin{eqnarray}
\frac{{B^2 }}{{8\pi }} = \frac{{Q\left( {{\rm{H}}^0 } \right)\left\langle {h\nu } \right\rangle }}{{4\pi R_{\rm{H}}^2 c}}
\label{eqn:balance}
\end{eqnarray}
\noindent
where standard notation is used (Osterbrock \& Ferland 2006).
Although this picture must be an oversimplification
it does account for the observed geometry and is physically plausible.

Guided by such considerations it should be possible to create an
underlying theory for why PDRs (the H$^0$ region) have their observed properties.
There are correlations between the mass of the molecular cloud (a fundamental
quantity), the ambient magnetic field in the cloud, and turbulent velocities
(Myers \& Goodman 1998).
The total luminosity of a star cluster is set by the mass of the
most massive star.
Equation 1.2 relates parameters such as $Q(\rm{H}^0)$ and $R_{\rm{H}}$, which can
be reposed as the $G_0$ PDR radiation parameter.
Similar relations can be derived for other quantities.

The magnetic field clearly plays a major role in establishing the geometry.
It is crucial that more measurements of fields near regions
of active star formation be made.
The measurements are difficult but the lessons learned are invaluable.

I thank the NSF (AST 0607028), NASA (NNG05GD81G), STScI
(HST-AR-10653) and the Spitzer Science Center (20343) for support.



\begin{thebibliography}{99}

\bibitem[\protect\citeauthoryear{Abel et al}{2004}]{AbelEtal06}
Abel, N. P., Brogan, C. L.,  Ferland, G. J., O'Dell, C. R., Shaw, G., \& Troland, T. H. 2004, ApJ, 609, 247-260

\bibitem[\protect\citeauthoryear{Abel et al}{2006}]{AbelEtal06}
Abel, N. P. Ferland, G.J. O'Dell, C.R. Shaw, G. \& Troland, T.H. 2006,
ApJ, 644, 344

\bibitem[\protect\citeauthoryear{Baldwin et al}{1991}]{BaldwinEtal91}
Baldwin, J., Ferland, G. J., Martin, P. G., Corbin, M., Cota, S., Peterson, B. M.,
\& Slettebak, A. 1991, ApJ, 374, 580

\bibitem[\protect\citeauthoryear{BBrogan et al}{1999}]{BroganEtal91}
Brogan, C. L., Troland, T. H., Roberts, D. A., \& Crutcher, R. M. 1999, ApJ, 515, 304

\bibitem[\protect\citeauthoryear{Ferland}{2001}]{Ferland01}
Ferland, G.J., 2001, PASP, 113, 41

\bibitem[\protect\citeauthoryear{Henney et al}{2005}]{HenneyEtal05}
Henney, W.J., Arthur, S.J., Williams, R.J.R., \& Ferland, G.J. 2005, ApJ, 621, 328

\bibitem[\protect\citeauthoryear{Henney et al}{2007}]{HenneyEtal07}
Henney, W.J., Williams, R. Ferland, G. Shaw, G \& C. O'Dell, 2007 ApJL, Dec 20

\bibitem[\protect\citeauthoryear{Lequeux}{2004}]{Lequeux04}
Lequeux, J, 2004, The Interstellar Medium, Springer

\bibitem[\protect\citeauthoryear{Myers \& Goodman}{1998}]{MyersGoodman98}
Myers, P.C. \& Goodman, A.A. 1998, ApJ, 326, L27

\bibitem[\protect\citeauthoryear{Osterbrock \& Ferland}{2006}]{AGN3}
Osterbrock D., Ferland G.J., 2006,
Astrophysics of gaseous nebulae and active galactic nuclei, 2nd.~ed

\bibitem[\protect\citeauthoryear{Pellegrini et al}{2007}]{PellegriniEtal07}
Pellegrini, E.W. Baldwin, J.A. Brogan, C.L. Hanson, M.M. Abel, N.P. Ferland,G.J.   Nemala, N.B. Shaw, G. \& Troland, T.H. 2007, ApJ, 658, 1119 (P07)

\bibitem[\protect\citeauthoryear{Roshi}{2007}]{Roshi07}
Roshi, D. Anish, 2007, ApJ, 658L, 48

\bibitem[\protect\citeauthoryear{Troland et al.}{1987}]{TrolandEtAl89}
Troland, T. H., Heiles, C., \& Goss, W. M. 1989, ApJ, 337, 342

\end{thebibliography}
\end{document}